\documentclass[twocolumn,showpacs]{revtex4}

\usepackage{epsfig}
\usepackage{graphicx}%
\usepackage{dcolumn}
\usepackage{amsmath}

\makeatletter
\def\btt#1{\texttt{\@backslashchar#1}}%
\DeclareRobustCommand\bblash{\btt{\@backslashchar}}%
\makeatother

\begin{document}

\title{CMBR Constraint on a Modified Chaplygin Gas Model}

\author{Dao-jun Liu}\email{cufa3@shnu.edu.cn}
\author{Xin-zhou Li}\email{kychz@shnu.edu.cn}

\affiliation{ Shanghai United Center for Astrophysics(SUCA),
Shanghai Normal University, 100 Guilin Road, Shanghai
200234,China\\
and\\
 Division of Astrophysics, E-institute of Shanghai
Universities, Shanghai Normal University, 100 Guilin Road,
Shanghai 200234,China
}%

\date{\today}

\begin{abstract}
In this paper, a modified Chaplygin gas model of unifying dark
energy and dark matter with exotic equation of state
$p=B\rho-\frac{A}{\rho^{\alpha}}$ which can also explain the
recent accelerated expansion of the universe is investigated by
the means of constraining the location of the peak of the CMBR
spectrum. We find that the result of CMBR measurements does not
exclude the nonzero value of parameter $B$, but allows it in the
range $-0.35\lesssim B\lesssim0.025$.
\end{abstract}


\maketitle

\section{Introduction}

The observational results of the anisotropy of cosmic microwave
background radiation (CMBR) \cite{WMAP}, supernovae type Ia (SNIa)
\cite{newobservation} and Sloan Digital Sky Survey (SDSS)
\cite{SDSS} show that about seventy percent of the total energy in
the universe should be hidden as dark energy which accelerates the
expansion of the universe at present. Several candidate to
represent dark energy have been suggested and confronted with
observations: cosmological constant, quintessence with a single
field \cite{Peebles} or with $N$ coupled fields \cite{Li}, phantom
field with canonical \cite{Caldwell} or Born-Infeld type
Lagrangian \cite{Hao}, k-essence \cite{Picon} and generalized
Chaplygin gas (GCG) \cite{Bento1} which is stemmed from the
Chaplygin gas \cite{Kamenshchik}.

In the GCG approach, the conjecture that dark energy and dark
matter can be unified by using an exotic equation of state has
been investigated. Within this framework, the difficulties of GCG
associated to unphysical oscillations or blow-up in the matter
power spectrum can be circumvented \cite{Bento2}. Recently, Hao
and Li \cite{Hao2} extend the equation of state of GCG to $w<-1$
regime. Many phenomenological tests have been considered: high
precision CMBR \cite{Bento1}, SNIa data \cite{Makler}, and
gravitational lensing \cite{Silva}. GCG can be extended to
modified Chaplygin gas (MCG) model \cite{Benaoum,Chimento} which
can also describe the current accelerating expansion of the
universe.

In this paper, we study the cosmological constraints on the MCG
model. The detailed structure of the anisotropies the CMBR depends
upon two epochs in cosmology: the emission of the radiation,
\textit{i.e.} last scattering and today. Therefore, the CMBR can
serve as a test to distinguish models of dark energy. Contrast to
the calculation of CMBR spectra, the location of the peaks and
troughs can be estimated with much less detailed calculation if we
assume that there exist adiabatic initial conditions and the
universe is flat \cite{Doran}.

\section{Modified Chaplygin Gas Model}

The energy density $\rho$ and pressure $p$ of MCG are related by
the following equation of state
\begin{equation}\label{eqnstate}
p= B\rho-\frac{A}{\rho^{\alpha}},
\end{equation}
where $B$, $A$ and $\alpha$ are three arbitrary constants. It is
obvious that when $B$ is zero, the above equation corresponds to
the equation of state of GCG, whereas when $A=0$ it reduces to the
equation of state of barotropic fluid.

In the flat $3$-dimensional Friedmann-Robertson-Walker spacetime
with metric
\begin{equation}
ds^2=-dt^2+a^2(t)d\vec{x}^2,
\end{equation}
\noindent  the equation of conservation can be written as
\begin{equation}\label{eqnConserv}
\dot{\rho}+3H(\rho+p)=0,
\end{equation}
\noindent where dot denotes the derivative with respect to cosmic
time $t$. From Eqs.(\ref{eqnstate}) and (\ref{eqnConserv}), the
density $\rho$ can be expressed by
\begin{equation}\label{rho1}
\rho_{_{MCG}}(a)=\rho_0\left[B_s+(1-B_s)\left(\frac{a_0}{a}\right)^{3(1+B)(1+\alpha)}\right]^{\frac{1}{1+\alpha}}
\end{equation}
\noindent for $B\neq -1$, and when $B=-1$,
\begin{equation}\label{rho2}
\frac{\rho_{_{MCG}}(a)}{\rho_0}=\left[1+A_s\ln\left(\frac{a_0}{a}\right)\right]^{\frac{1}{1+\alpha}}.
\end{equation}
\noindent Here $\rho_0$ denotes the energy density of MCG at
$t=t_0$ and the integration constants $A_s$ and $B_s$ are defined
as
\begin{equation}
A_s=\frac{3A(1+\alpha)}{\rho_0^{1+\alpha}}
\end{equation}
and
\begin{equation}
B_s=\frac{A}{(1+B)\rho_0^{1+\alpha}}
\end{equation}
respectively. For the flat universe, the Friedmann equation reads
\begin{equation}
H^2\equiv\left(\frac{\dot{a}}{a}\right)^2=\frac{8\pi
G}{3}\left(\frac{\rho_{b,o}}{a^3}+\frac{\rho_{r,0}}{a^4}+\rho_{_{MCG}}\right),
\end{equation}
\noindent where $\rho_{b,0}$ and $\rho_{r,0}$ are the energy
density of baryons and radiation at $t=t_0$, and $\rho_{_{MCG}}$
is energy density of MCG represented in Eq.(\ref{rho1}) or
Eq.(\ref{rho2}). We do not include dark matter in that dark matter
is accounted for by MCG. Notice that $B_s$ must lie in the
interval $0\leq B_s \leq 1$ because otherwise $\rho_{_{MCG}}$
would be undefined for some time in the past. Therefore, we can
find that at early epoch the energy density behaves as matter
while at late time it behaves like a cosmological constant except
the MCG models with $B=-1$. From Eq.(\ref{rho2}), it is easy to
find that $\rho_{MCG}$ is only well-defined in the regime $a\geq
a_0e^{1/A_s}$ for the case $B=-1$. It is not difficult to find
that the MCG behaves as dust matter for $B_s=0$ and cosmological
constant for $B_s=1$. Unlike the GCG Model, the MCG model does not
correspond to a $\Lambda$-CDM model when $\alpha=-1$, but does
when $\alpha=\frac{-B}{1+B}$ .

Define the new variables $\Omega_{b,0}\equiv
\frac{\rho_{b,0}}{\rho_{c,0}}$ and $\Omega_{r,0}\equiv
\frac{\rho_{r,0}}{\rho_{c,0}}$, where
$\rho_{c,0}=\frac{3H_0^2}{8\pi G}$ is the critical density of the
universe and $H_0$ is the Hubble parameter at $t=t_0$.  Using the
fact that $H= \frac{1}{a^2}\frac{da}{d\tau}$ and taking scale
factor $a_0=1$, we obtain
\begin{equation}
\frac{da}{d\tau}=H_0 X(a),
\end{equation}
\noindent where
\begin{widetext}
\begin{equation}
X(a)=\sqrt{\Omega_{r,0}+\Omega_{b,0}a+a^4(1-\Omega_{r,0}-\Omega_{b,0})\left[B_s+\frac{1-B_s}{a^{3(1+A)(\alpha+1)}}\right]^{\frac{1}{1+\alpha}}},
\quad\text{for} \quad B\neq -1,
\end{equation}
\noindent  and
\begin{equation}
X(a)=\sqrt{\Omega_{r,0}+\Omega_{b,0}a+a^4(1-\Omega_{r,0}-\Omega_{b,0})\left(1+A_s\ln(a)\right)^{\frac{1}{1+\alpha}}},
\quad\text{for}\quad B= -1.
\end{equation}
\end{widetext}

\section{The location of the CMBR spectrum peaks}
The CMBR peaks arise from acoustic oscillations of the primeval
plasma just before the universe becomes transparent. The angular
momentum scale of the oscillations is set by the acoustic scale
$\l_A$ which for a flat Universe is given by
\begin{equation}
l_A=\pi\frac{\tau_0-\tau_{ls}}{\bar{c_s}\tau_{ls}}.
\end{equation}

\noindent Here  $\tau_0$ and $\tau_{ls}$ are the conformal time
today and at last scattering which are equal to the particle
horizons and $\tau=\int\frac{dt}{a(t)}$ with cosmological scale
factor $a$; while $\bar{c_{s}}$ is the average sound speed before
decoupling,
\begin{equation}
\bar{c_{s}}=\frac{\int_0^{\tau_{ls}}c_s(\tau)d\tau}{\tau_{ls}},
\end{equation}
\noindent where $c_s$ is satisfied by
\begin{equation}
c_s=\left[3+\frac{9}{4}\frac{\rho_b(\tau)}{\rho_{\gamma}(\tau)}\right]^{-1/2},
\end{equation}
\noindent with $\rho_b/\rho_{\gamma}$ the ratio of baryon to
photon energy density. So that,
\begin{equation}
l_A=\frac{\pi}{\bar{c}_s}\left[\frac{\int_0^1\frac{da}{X(a)}}{\int_0^{a_{ls}}\frac{da}{X(a)}}-1\right].
\end{equation}
\noindent where $a_{ls}$ is the scale factor at the last
scattering time, for which we use the fitting formula developed in
Ref.\cite{Hu2},
\begin{equation}
a_{ls}^{-1}+1=1048[1+0.00124(\Omega_{b,0}h^2)^{-0.738}][1+g_1(\Omega_{m,0}h^2)^{g_2}],
\end{equation}
\noindent where
\begin{equation}
g_1=0.0783(\Omega_{b,0}h^2)^{-0.238}[1+39.5(\Omega_{b,0}h^2)^{0.763}]^{-1}
\end{equation}
\begin{equation}
g_2=0.56[1+21.1(\Omega_{b,0}h^2)^{1.81}]^{-1}.
\end{equation}
 The location of the $m$-th peak can be approximated by
\cite{Hu}
\begin{equation}
l_m=l_A(m-\phi_m)
\end{equation}
\noindent where $\phi_m$ is the $m$-th peak shift. The Analytical
relationships between the cosmological parameters and the peak
shifts are not available, but one can use fitting formulae
describing their dependence on these parameters \cite{Bento1}.
Particularly, for the spectral index of scalar perturbations
$n=1$, the shifts of the first peak can be expressed approximately
by
\begin{equation}
\phi_1=0.267\left(\frac{r_{ls}}{0.3}\right)^{0.1},
\end{equation}
\noindent where the ratio of radiation to matter at last
scattering $r_{ls}$ is fitted by
\begin{equation}
r_{ls}\equiv\frac{\rho_{r,ls}}{\rho_{m,ls}}=\frac{0.042}{10^3\Omega_{m,0}h^2}\left(\frac{1}{a_{ls}}-1\right).
\end{equation}
According to the WMAP measurements of the CMBR temperature angular
power spectrum,  the first peak is located at
\begin{equation}\label{peak1}
l_1=220.1\pm 0.8,
\end{equation}
\noindent with $1\sigma$ uncertainty.

In FIG.1, we plot the constraint on the three parameters $B_s$,
$\alpha$ and $B$ of the MCG model($B\neq -1$) corresponding to the
bounds on the first peak of the CMBR power spectrum
(Eq.(\ref{peak1})) for spectral index $n=1$.

\begin{figure}
\epsfig{file=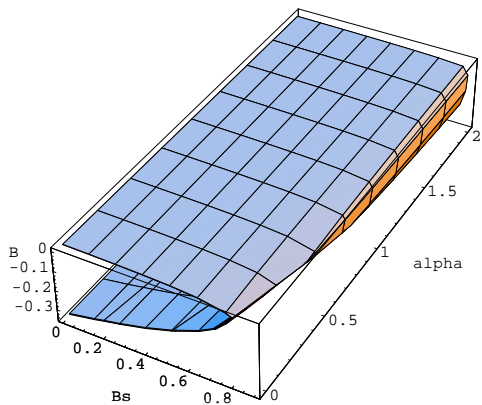,height=2.1in,width=2.5in}
\caption{constraints on the three parameters $B_s$, $\alpha$ and
$B$ of the MCG model($B\neq -1$) corresponding to the bounds on
the first peak of the CMBR power spectrum for spectral index
$n=1$, $h =0.71$, $\Omega_{b,0}=0.046$ and $\Omega_{r,0}=9.89
\times 10^{-5}$ .}
\end{figure}

\begin{figure}
\epsfig{file=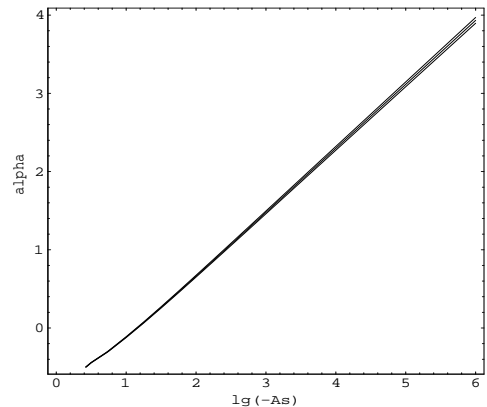,height=2.1in,width=2.5in}
\caption{Constraint on the parameters $A_s$ and $\alpha$ of the
MCG model with $B = -1$ corresponding to the bounds on the first
peak of the CMBR power spectrum for spectral index $n=1$, $h
=0.71$, $\Omega_{b,0}=0.046$ and $\Omega_{r,0}=9.89 \times
10^{-5}$ .}
\end{figure}

The constraint on the parameters $A_s$ and $\alpha$ of the MCG
model with $B = -1$ are plotted in FIG.2 corresponding to the same
bounds as that of FIG.1.

Modified Chaplygin gas Model of unifying dark energy and dark
matter with exotic equation of state can explain the current
accelerating expansion of the universe.  By investigating the
constraint from the location of the peak of the CMBR spectrum, we
find that the result of CMBR measurements does not exclude the
nonzero value of parameter $B$, but allows it in the range
$-0.35\lesssim B\lesssim 0.025$.

It is worth noting that an interesting model that investigated in
Ref. \cite{Chimento} is included in (\ref{eqnstate}). As a special
case, it requires that $B$ is related to $\alpha$ via
$B=-\frac{\alpha}{1+\alpha}$. Obviously, $B$ can not be equal to
$-1$ for finite value of $\alpha$. Therefore, it is available for
this case to calculate the energy density by using
Eq.(\ref{rho1}). In FIG.3, we plot the constraint on the
parameters $B_s$ and $\alpha$ of this kind of MCG model
($B=-\frac{\alpha}{1+\alpha}$) corresponding to the bounds on the
first peak of the CMBR power spectrum for spectral index $n=1$.
From Fig.3, it is easy to find that for reasonable value of $B_s$,
the value of $\alpha$ is constrained in a range of $-0.021\lesssim
\alpha \lesssim 0.54$. This equivalently means the parameter $B$
is allowed in the range $-0.35\lesssim B\lesssim 0.021$, which is
covered in general case.

\begin{figure}
\epsfig{file=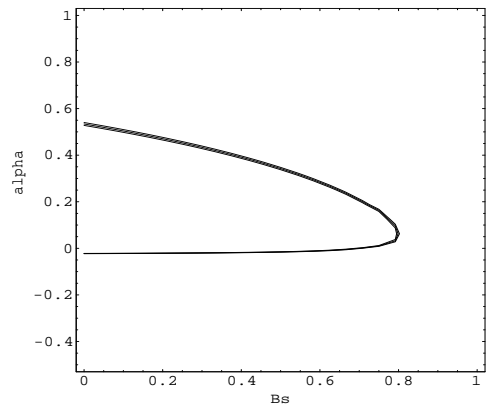,height=2.1in,width=2.5in}
\caption{Constraint on the parameters $B_s$ and $\alpha$ of the
MCG model with $B = -\frac{\alpha}{1+\alpha}$ corresponding to the
bounds on the first peak of the CMBR power spectrum for spectral
index $n=1$, $h =0.71$, $\Omega_{b,0}=0.046$ and
$\Omega_{r,0}=9.89 \times 10^{-5}$ .}
\end{figure}

\section*{Acknowledgments}

This work is supported by Shanghai Municipal Education Commission
(No.04DC28) and National Natural Science Foundation of China under
Grant No. 10473007.

\end{document}